\documentclass[conference]{IEEEtran}
\usepackage{ifpdf}

\usepackage{amssymb}
\usepackage{cite}

\ifCLASSINFOpdf
   \usepackage{graphicx}
   \graphicspath{{Figures/}}   
   \graphicspath{{../pdf/}{../jpeg/}{../eps/}}
   \DeclareGraphicsExtensions{.pdf,.jpeg,.png}
\else
   \usepackage[dvips]{graphicx}
   \graphicspath{{../eps/}}
   \DeclareGraphicsExtensions{.eps}
\fi
\usepackage[cmex10]{amsmath}
\usepackage{multirow}

\begin{document}
%
\title{Reliability Modeling, Analysis and Prediction of Wireless Mobile Communications}

\author{\IEEEauthorblockN{Raja Sattiraju}
\IEEEauthorblockA{Chair for Wireless Communication\\and Navigation\\
University of Kaiserslautern\\
Kaiserslautern, Germany\\
Email:sattiraju@eit.uni-kl.de}
\and
\IEEEauthorblockN{Hans D. Schotten}
\IEEEauthorblockA{Chair for Wireless Communication\\and Navigation\\
University of Kaiserslautern\\
Kaiserslautern, Germany\\
Email:schotten@eit.uni-kl.de}}


%


\maketitle

\begin{abstract}
The future Fifth Generation (5G) mobile cellular networks that are currently in research phase today enable broad range of services/applications beyond classical mobile communications. One key enabler for Ultra-Reliable services to be integrated into mobile networks is the \textit{Reliability} of transmission success of a given data packet. This is harder mainly owing to the time-dependent effective link qualities of the communicating devices. However, \textit{successful} indication of the availability of the instantaneous link quality (e.g., by the device) would allow opportunistic access of ultra reliable services/applications when the link conditions are fair enough. This paper introduces a framework for modeling, predicting and analyzing the theoretical reliability of the wireless link based on factors such as fading, mobility, interference etc. The analysis and prediction is based on the part stress method\cite{Birolini2010}  by assuming time dependent factors as elements/components and their respective Transmission Times To Failure (TTTF). The proposed framework also supports other reliability analysis techniques such as Fault Tree Analysis\cite{5222114} and Accelerated testing\cite{Nelson2009} of wireless systems and to \textit{improve} the components.
\end{abstract}


%
\IEEEpeerreviewmaketitle

\section{Introduction}

{\let\thefootnote\relax\footnote{This is a preprint, the full paper is published in Proceedings of 79th IEEE Vehicular Technology Conference(IEEE VTC Spring 2014), \copyright 2014 IEEE. Personal use of this material is permitted. However, permission to use this material for any other purposes must be obtained from the IEEE by sending a request to pubs-permissions@ieee.org.}}

The exponential increase in the demand for mobile and wireless connectivity is witnessed by the increasing number of new services and applications, many of which require stringent reliability requirements. Examples of such services/application are Vehicle-to-Vehicle(V2V), Vehicle-to-Infrastructure(V2I), and other industrial applications(Sensor Networks)with varying design criteria which are typically specified in terms of Availability, rather than Reliability\cite{Ayers2012}. Analysis of these systems requires specialized application of reliability engineering theory and its principles\cite{Rak2013} with performance expectations ranging from high to extreme\cite{Ayers2012}.\  

The wireless applications of automotive industry have been classified into Safety, Convenience and Commercial categories\cite{Bai} from a value or consumer benefit perspective with varying reliability requirements. From the implementation point of view, reliability analysis of these supported applications/services on par with the industrial standards would for example, allow safe integration of the service/application as a component.\

EU Projects (METIS)\cite{metis2020} have already started working on designing new wireless interface solutions for supporting  varying reliability requirements. From the implementation point of view, Reliability is defined as the probability that a packet is delivered error-free before the deadline expires. This is inline with the definition of reliability from the engineering point of view which says that "Reliability is the characteristic of the item, expressed by the probability that it will perform its required function under given conditions for a stated time interval"\cite{Birolini2010}. The required function in this case is the successful transmission of the packet (link reliability) which is based on the following reliability factors (operating conditions)\cite{METIS2013}.
\begin{itemize}
	\item Decreased power of the useful signal due to fading, mobility, etc
	\item Increased uncontrollable interference from licensed or unlicensed devices
	\item Resource depletion due to competing devices such as in a random access, downlink congestion, etc. 
	\item Protocol reliability mismatch captures the fact that the protocols operate as expected in a region where the control messages are reliably transferred, but the control messages are commonly not adaptive to seriously worsened communication conditions
	\begin{itemize}
          \item e.g.supporting high throughput, but not the capability to send an alarm message with low latency          
    \end{itemize}
\end{itemize}

The purpose of this paper is to provide a framework for modeling, analyzing and prediction of the wireless "link reliability" using the methods and techniques presented in Reliability Engineering\cite{Birolini2010} starting with the first reliability factor presented above which deals with modeling of channel fading.

\subsection{Wireless Channel Fading} 
In digital communication systems, the most frequently used channel model is the Additive White Gaussian Noise (AWGN) channel. However, this model is too simplistic and cannot support complex spatial and temporal variations\cite{Arsal2008} due to more basic factors such as pathloss, shadowing and multi-path propagation affecting the radio propagation.Each of this phenomenon is caused by a different underlying physical principle (Reflection, Refraction, Diffraction and Scattering of electromagnetic wave) and must be considered when designing and evaluating wireless communication systems.\\
Among the three independent phenomena, only path loss is deterministic which depends only on the distance between Tx and Rx. Pathloss is more relevant on larger time scales like seconds/minutes, since the distance between the Tx and the Rx does not significantly change on smaller time scale\cite{Arsal2008}. On smaller time scales, its effect is visible for high mobile communications (mobility). On the other hand, shadowing and multi-path fading are not deterministic and have a stochastic nature\cite{Poikonen2009}. Shadowing occurs due to varying terrain conditions due to the presence of obstacles. Fading is caused by multi-path propagation environment and leads to significant attenuation changes within smaller time scales such as milliseconds or even microseconds.

\subsection{Organization of the Paper}
The paper is organized into V sections. Section I already dealt with the introduction of link reliability and the affecting factors. Section II presents the system model and deals with reliability modeling of the wireless channel  using Transmission Time To Failure (TTTF) and the state variable X(t) of the wireless system $S_w$. Section III provides the Reliability Analsysis techniques and the selection of suitable techniques for analysing wireless systems. Section IV considers an example method to analyse the reliability of the considered system and presents some initial results. Section V concludes the paper and provides a brief note of the future works planned.

\section{Reliability modeling of wireless channel}
In this section, we deal with the reliability modeling of a wireless channel considering the three physical phenomena presented in Section 1 - Pathloss, Shadowing and Multi-path fading.
\subsection*{System Model}
For the purpose of simplification, let us consider a receiver Rx within a certain distance from the transmitter Tx. Let the state variable X(t) represent the operational state of the transmission in a time slotted model at time $t$ where
 \[ \text{X}(t) = \left\{ 
  \begin{array}{l l}
    1 & \quad \text{if the transmission is operational at time $t$}\\
    0 & \quad \text{if the transmission failed at time $t$}
  \end{array} \right.\]
The state variable X(t) is the fundamental unit of reliability analysis. Although this discussion is limited to the "functional" and "failed" states, the analysis can be expanded to allow X(t) to assume any number of different states (for e.g., providing differentiated levels of reliability)  
 
The Reliability/Survival function R(t) is a mathematical expression analytically relating the probability of successful transmission to time and can be described by the state variable $X(t)$ and the Transmission Time to Failure (TTTF) which is the measure of of the amount of time elapsed before the transmission fails. TTTF can either be a discrete or a continuous valued function.

Let the TTTF of any given transmission be given by a random variable $t$. We can thus write the probability that the transmission time to failure T is greater than t=0 and less than a time t (this is also called as the CDF F(t) on the interval [0,t)) as
\[f(t)=Pr(T\leqslant t)\;  for\;  [0,t)]]\]

The Cumulative Distribution Function(CDF), can be derived from the PDF by evaluating the relationship
\[F(t)=\int_{0}^{t}f(t)dt\; for\: all\: t\geq 0\]

where f(t) is the PDF of the transmission time to failure. Conceptually, the PDF represents a histogram function of time for which f(t) represents the relative frequency of occurance of TTTF events.The reliability of the transmission is the probability that an item does not fail for an interval (0,t]. For this reason, the reliability function R(t) is also referred to as the survivor function since the transmission “survives” for the time t. Mathematically, we can write the reliability/survivor function R(t) as
\[R(t)=1-F(t)\;\;\; \;   t\geq 0\]
\
The Probability Density Function (PDF) describes the probability that the transmission fails over time and is also referred to as Life distribution of the item. Failure rate ($\lambda$) is a very important metric for reliability analysis which gives the conditional probability that the transmission operating failure-free until $t$ ($R(T=t)=1$) also survives the period $\Delta t$
\[\lambda(t)=\frac{R(t + \Delta t)}{R(t)}
\]
Thus, the probability that the transmission fails within $\Delta t$ is
\[1-\frac{R(t + \Delta t)}{R(t)}=\frac{1-F(t)-(1-F(t + \Delta t))}{1-F(t)}\]
As the given probability for short time intervals $\Delta t$ is proportional to $t$, hence we divide the term by $\Delta t$ and determine the boundary value when $\Delta t$ approximates zero.
\[
\begin{split}
&\lim_{\Delta t \rightarrow 0}\; \frac{1}{\Delta t}\; \frac{F(t+\Delta t)-F(t)}{1-F(t)}\\
&=\frac{1}{R(t)}\; \lim_{\Delta t\rightarrow 0}\; \frac{F(t +\Delta t)-F(t)}{\Delta t}=\frac{f(t)}{R(t)}=\lambda(t)
\end{split}
\] 
Thus, the probability of the transmission, that is operational at time $t$ fails within the (short) time interval $\Delta t$, is approximately $\Delta t* \lambda(t)$. \\
The failure rate ($\lambda$) is also known as Hazard function and applies in particular to non-repairable systems. For repairable systems, failure rate translates into Availability which will be investigated in further work. Hence, to model the transmission reliability, the most important parameters are the TTTF and the PDF (life distributions) of the factors involved. The life distributions of pathloss, shadowing and multi-path fading can be accordingly derived from their emperical/statistical models. 
\subsection*{Life Distributions of Fading Components}
Pathloss (large scale pathloss) is deterministic in nature for low mobility due to spatial variations of the received signal. However for high mobility scenarios the temporal variations (due to channel coherence time) make the fading phenomenon more in-deterministic\cite{Arsal2008}. In simplistic scenarios, pathloss can be generally modeled by an exponential distribution with varying values of path loss exponent\cite{5044933} as shown in Fig.1\\
\begin{figure}[h]
\minipage{0.24\textwidth}
  \includegraphics[width=\linewidth]{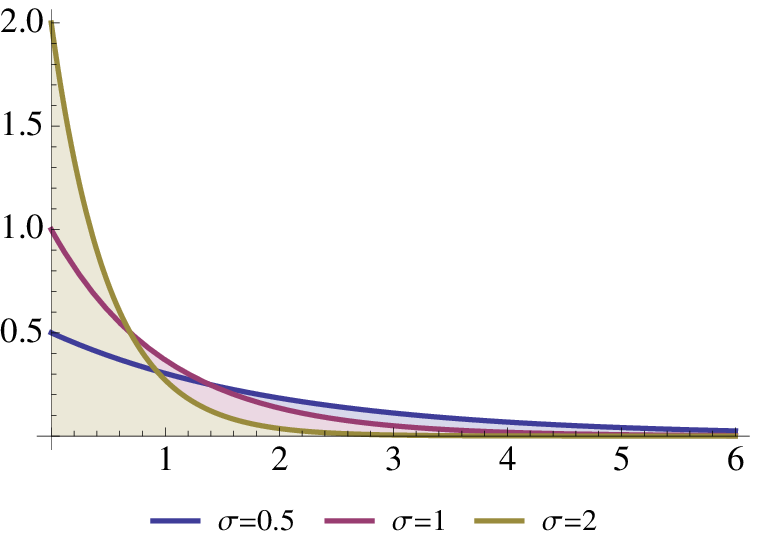}
  
\endminipage\hfill
\minipage{0.24\textwidth}
  \includegraphics[width=\linewidth]{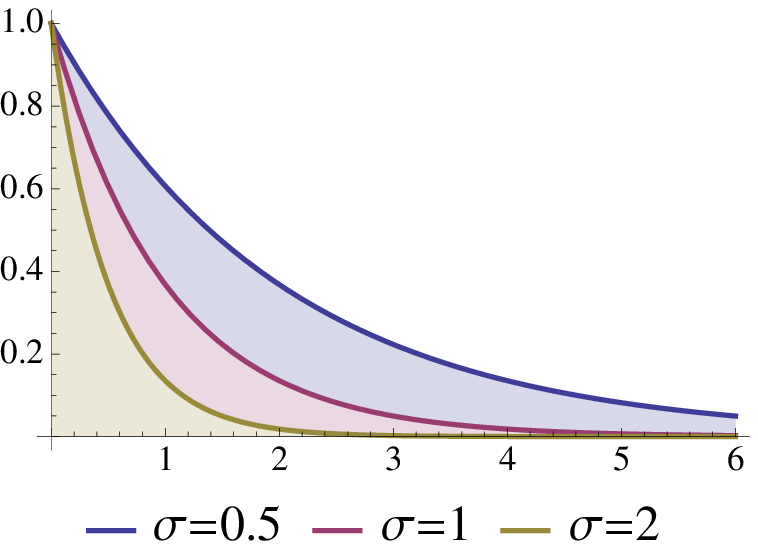}
  
\endminipage\hfill
\caption{$f(t)$ and $R(t)$ for Large-scale pathloss}
\end{figure}

Shadowing occurs due to the obstruction of the line of propagation. Thus, for a given fixed distance, frequency and transmission power, the received signal is not deterministic anymore and can be only modeled stochastically as a function of distance. Experimental results show that the shadow fading can be fairly accurately modeled as a log-normal random variable\cite{Reudink1972}\cite{okumura1968field}\cite{Gudmundson1991}. The theoretical basis is that in a propagation environment, different signals suffer reflection and diffraction as they traverse the propagation medium. If the total loss of the signal at Rx is expressed in dB, then the extra loss due to shadowing in each signal path corresponds to subtracting a random loss from this path loss value. Since the different propagation paths are independent, the sum of all the dB losses for a large number of propagation paths converges to a normal distributed random variable (central limit theorem). In natural units, that becomes a log-normal distribution which also represents the life distribution (Fig.2) of the shadowing component.\\
\begin{figure}[h]
\minipage{0.24\textwidth}
  \includegraphics[width=\linewidth]{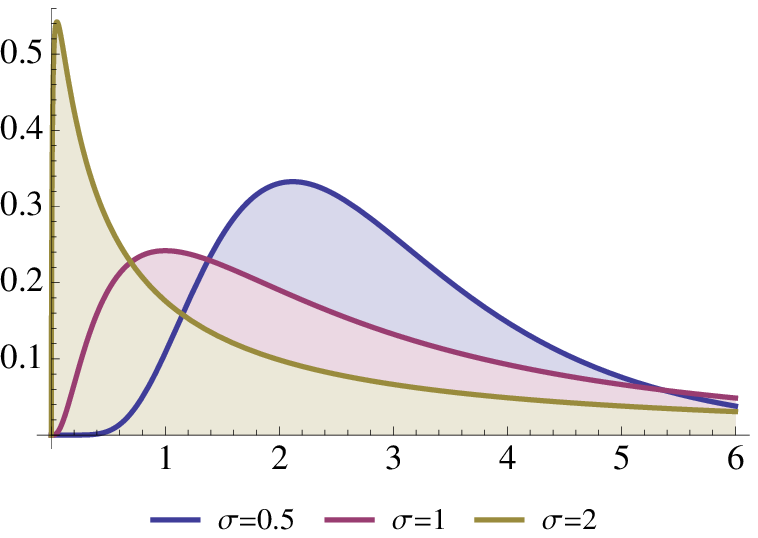}
  
\endminipage\hfill
\minipage{0.24\textwidth}
  \includegraphics[width=\linewidth]{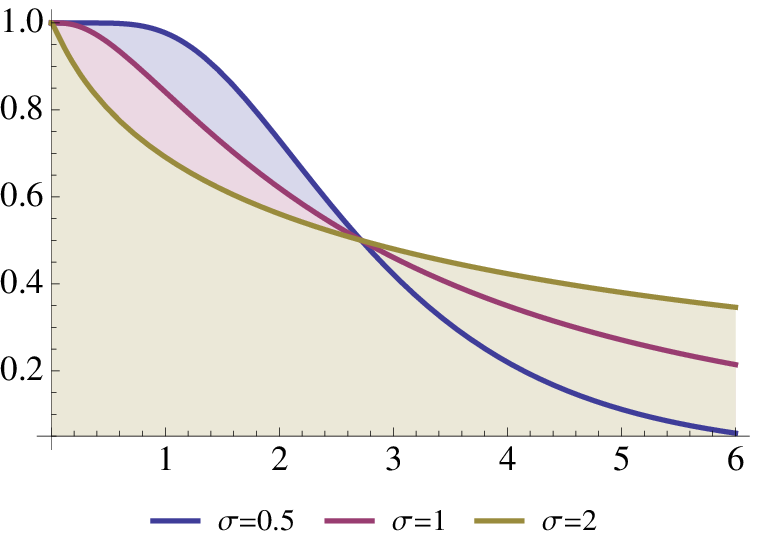}
  
\endminipage\hfill
\caption{$f(t)$ and $R(t)$ for Shadowing}
\end{figure}

Multi-Path fading occurs due to the constructive-destructuve superimposition of arriving multiple signals at the Rx. The destructuve interference is generally modeled by Rayleigh (Fig.3) or Rician distribution. However, complex statistical models such as Clarke and Gans Fading model\cite{clarke1968statistical},Saleh and Valenzuela Indoor Statistical Model\cite{saleh1987statistical},SIRCIM and SMRCIM Indoor and Outdoor Statistical Models\cite{rappaport1991statistical} etc. try to model the multi-path fading accurately by real world measurements. More accurate models are presently investigated by using computationally extensive ray tracing techniques and terrain data.\\
\begin{figure}[h]
\minipage{0.24\textwidth}
  \includegraphics[width=\linewidth]{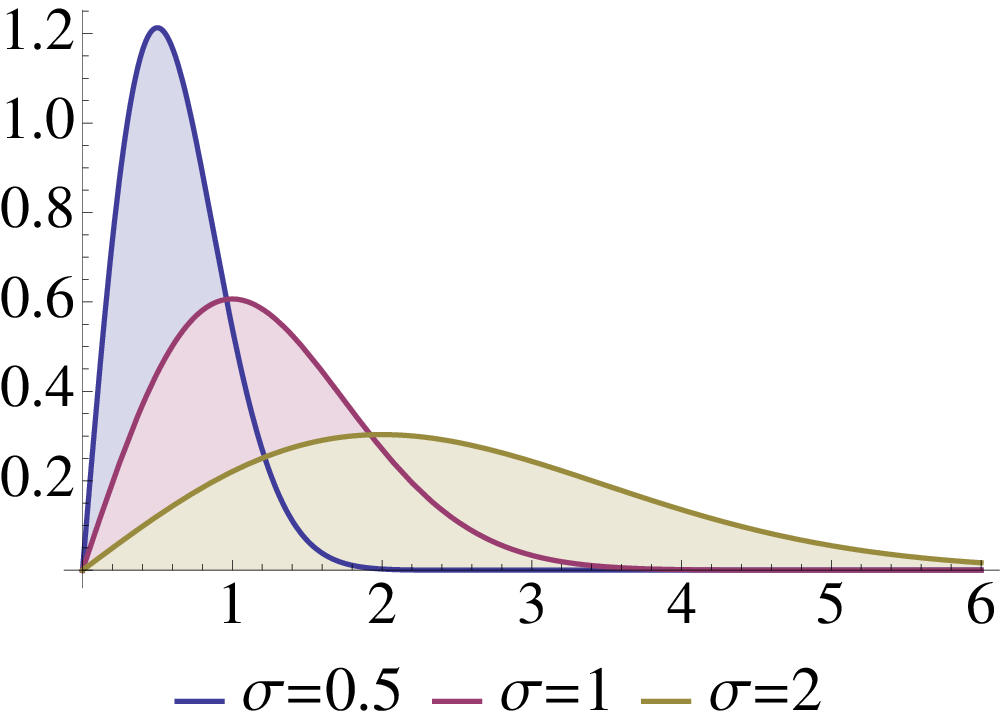}
  
\endminipage\hfill
\minipage{0.24\textwidth}
  \includegraphics[width=\linewidth]{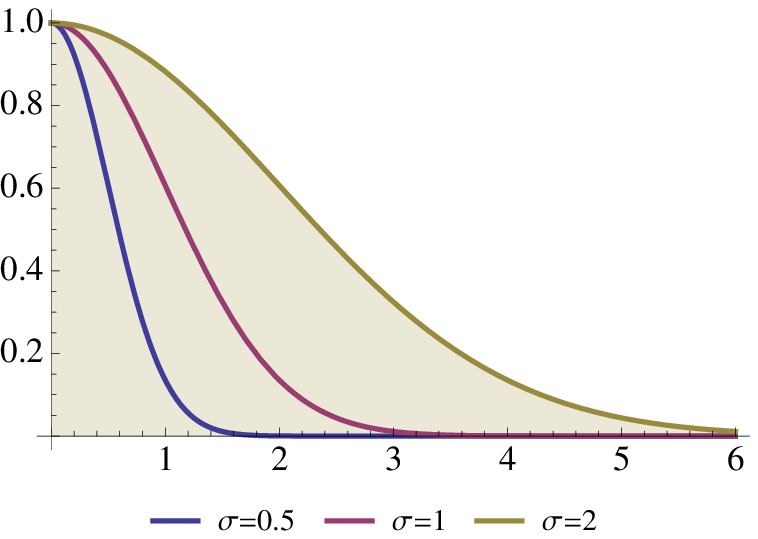}
  
\endminipage\hfill
\caption{$f(t)$ and $R(t)$ for multi-path fading}
\end{figure}

\section{Reliability Analysis}
An important part of the reliability analysis deals with the failure rate and failure mode investigation and calculating the \textit{predicted reliability}, ie., that reliability which can be calculated from the structure of the system and the reliability of its elements\cite{Birolini2010}.\\

On the other hand, the \textit{true reliability} of the system can only be determined by reliability tests by evaluating the performance of the wireless transmission in real-time and collecting the Bit Error Rate (BER) or Packet Error Rate (PER) statistics of the transmitted packet. One approach for statistical reliability analysis is based on the \textit{part stress method}\cite{Birolini2010} (refer to Fig.4) to check the validity of the assumed failure modes, and a verification of the adherence to design guidelines for reliability in a preliminary design review.

\begin{figure}[h!]
\centering
\includegraphics[width=0.45\textwidth]{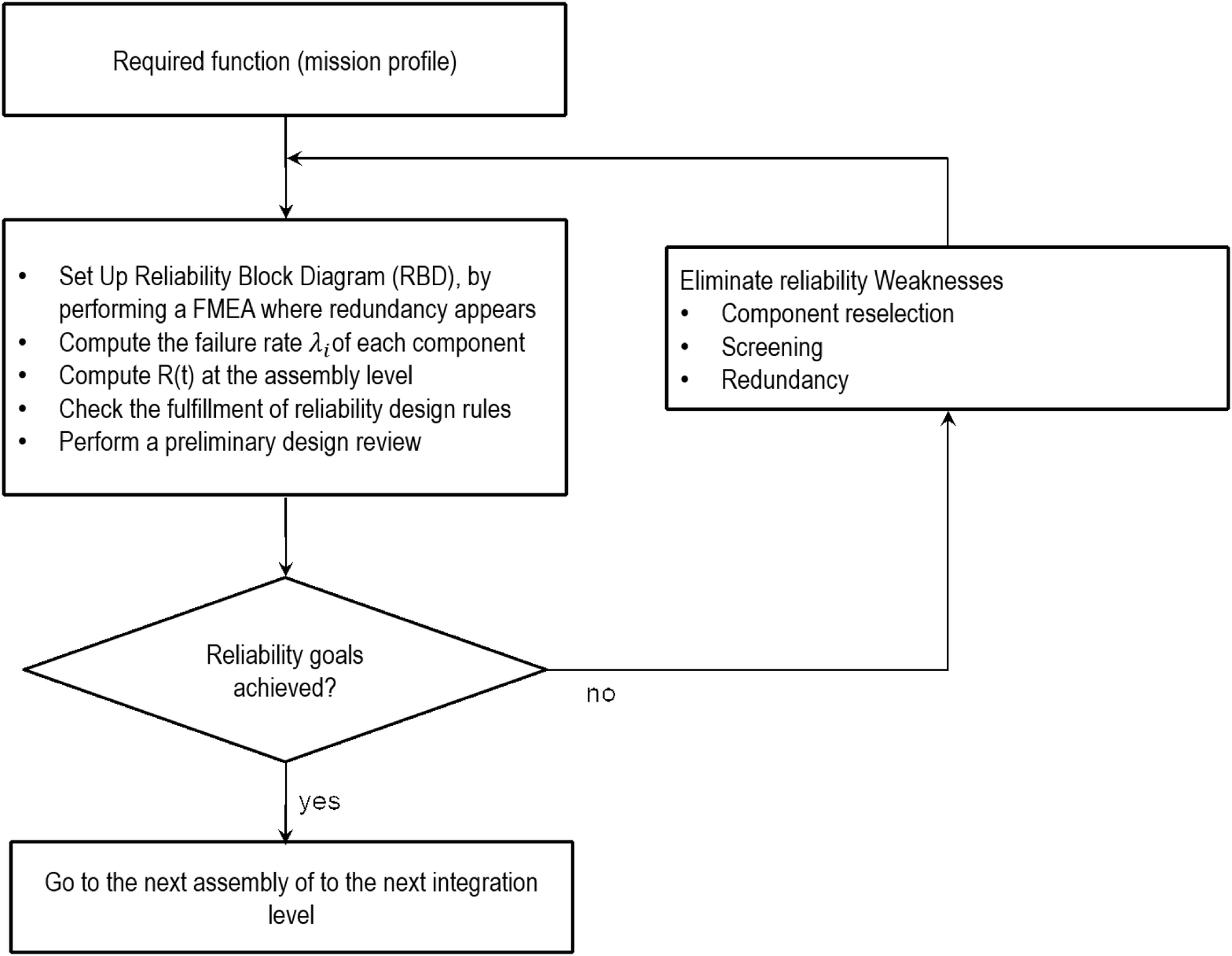}
\caption{Reliability Analysis flowchart}
\label{fig_sim}
\end{figure} 

\subsection{Required function}
The required function here is the successful transmission of the packet of size $n$ under a given deadline. The deadline and the packet size will be specified by the service/application itself via an Availability Request (AR) and Availability Response as described in \cite{Sattiraju2013}. Such an AR from the service/application will not only help indicate the reliability of the instantaneous link quality but also to provide differentiated grades of service with respect to required latencies. Ultra low latency can be achieved for example, by limiting the packet length (using short packet lengths) and limiting the control information flow (minimizing protocol handshakes)\cite{METIS2013}.

\subsection{Derivation of Reliability Block Diagram and calculation of System Reliability $R_w$}
The Reliability Block Diagram (RBD) is an event diagram. It answers the following question: Which elements of the system under consideration are necessary for the fulfillment of the required function and which can fail without affecting it? Setting up a RBD involves, at first, partitioning the item into elements with clearly defined tasks\cite{Birolini2010}. The elements which are necessary for the required function are connected in series, while elements which can fail with no effect on the required function (redundancy) are connected in parallal. The ordering of the series elements in the RBD can be arbitrary. RBD can only consider two states (e.g., good or bad) and one failure mode (e.g., opens or shorts) for each element. RBD also differs from the functional block diagram since one or more elements can appear more than once in RBD.\
The wireless transmission system $S_w$ is assumed to be a series system with no redundancy since all the elements considered (pathloss, shadowing and multi-path fading) must work in order to fulfill the required function (successful transmission). However, in reality factors such as retransmissions, constructive multi-path fading etc. contribute to the transmission redundancy and are not considered in our simplified approach. The System Reliability $R_w$ can be calculated as the joint probability distribution (reliability distribution) as
\[R_{w}=R_{pathloss}\wedge R_{shadowing}\wedge R_{multipath_fading}\]

\subsection{Preliminary Design Review}
The design review consists of analyzing the reliability functions of individual elements and identifyng the elements of improvement. There are many importance metrics for identifying and finding weaknesses in a system\cite{Hwang2001}. Some of them are 
\subsubsection*{Birnbaum Importance}
Also known as the reliability importance, it is the improvement in the reliability that would be gained by replacing a failed element $i$ with a perfect element $i$. The Birnbaum importance at time $t$ for component $i$ is given by $p_i(t)-q_i(t)$ where $p_i(t)$ is the probability that the system is working given that the $i^{th}$ component is perfect and $q_,i(t)$ is the probability that the system is working given that the $i^{th}$ component has failed.

\subsubsection*{Improvement Importance}
Also known as improvement potential, the improvement importance of a component $i$ is the increase of the Survival Function if $i$ is replaced with a perfect element. The improvement importance at time $t$ for component $i$ is given by $p_i(t)-p(t)$, where $p_i(t)$ is the probability that the system is working, given that the $i^{th}$ component never fails, and $p(t)$ is the probability that the system is working.\\
Other Importance metrics like Barlow Proschan Importance, Criticality Failure Importance and Criticality Success Importance\cite{barlow1975reliability} can also be calculated.

\subsection{Eliminate Reliability Weaknesses}
In the context of wireless systems, reliability weaknesses result from the reliability factors contributing to the failure of transmission. However, techniques like link adaptation (e.g.,via power control), restransmissions, MCS schemes etc. can be used to eliminate the reliability weaknesses. 

\section{Simulation Results}
The Key Performance Indicators (KPI's) for calculating the system reliability $R_w$ of the system $S_w$ are the survival/reliability function $R(t)$ and the hazard function/ Failure rate $\lambda(t)$. These are calculated via the individual life distributions of the factors. Joint probability distributions are then used to derive the system reliability $R_w$.

\subsection{Life Distributions}
The failure/life distribution assumptions of pathloss, shadowing and multipath fading are outlined in  Table I.
\begin{table}[h]
\renewcommand{\arraystretch}{1.3}
\caption{Life Distributions of the Components}
\label{table_example}
\centering
\begin{tabular}{|c||c|c|c| }
\hline
Element &  Distribution & Mean & Variance\\
\hline
Pathloss & Exponential & 1 & 1\\
\hline
Shadowing & Log-Normal & $e^3$ & $e^6(-1+e^4)$\\
\hline
Multipath Fading & Rayleigh & $\sqrt{2\pi }$ & $4(2-\frac{\pi }{2})$\\
\hline
\end{tabular}
\end{table}\
R(t) and $\lambda(t)$ for each of the component can be recursively calculated as shown in Section II.

%
%

\subsection{System Reliability $R_w$}
Since $S_w$ is assumed to be a series system (the failure of any of the element results in the failure of the transmission), the total system reliability $R_w$ can be calculated as the product of the component reliability\cite{Birolini2010}
\[R_{W}=R_{pathloss}*R_{shadowing}*R_{multipath}\]

\begin{figure}[h!]
\centering
\includegraphics[width=0.5\textwidth]{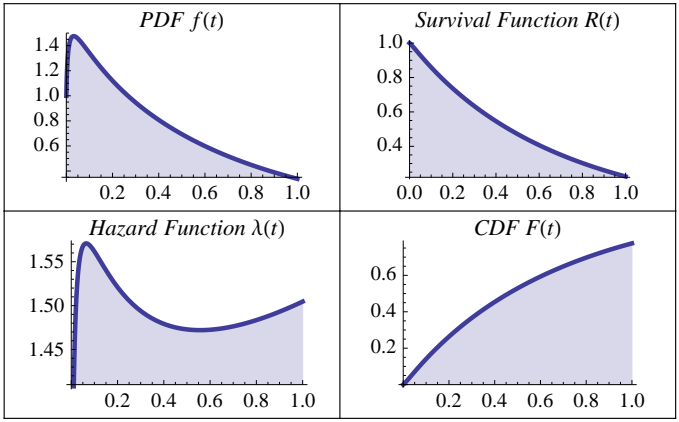}
\caption{System Reliability($R_w$) KPI's}
\end{figure} 

Fig.5 shows the survival and hazard functions of $S_w$ considering pathloss, shadowing and multipath fading. The TTTF (per unit time) can be calculated as the mean of the survival function and is found to be $0.65$. The Probability of transmission failure before time $t$ is also calculated per unit time and the results are given in Table II.\

\begin{figure}[!h]
\minipage{0.23\textwidth}
  \includegraphics[width=\linewidth]{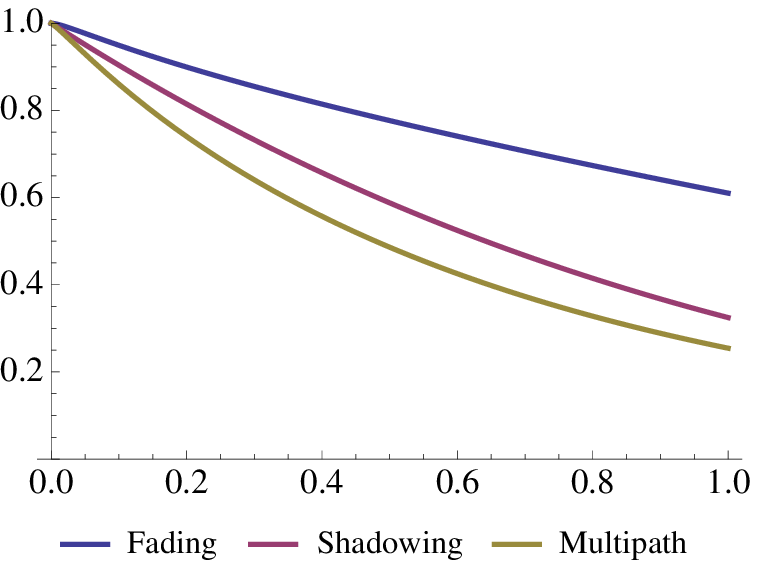}
  
\endminipage\hfill
\minipage{0.23\textwidth}
  \includegraphics[width=\linewidth]{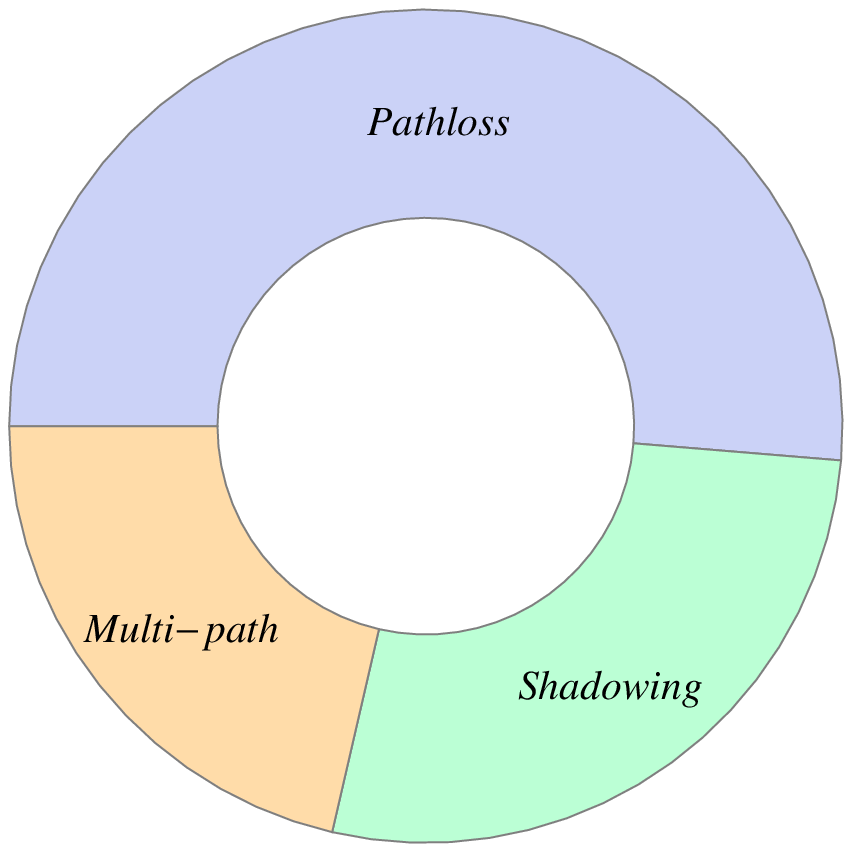}
  
\endminipage\hfill
\caption{Birnbaum Importance for $S_w$}
\end{figure}

The Birnbaum importance metric of $S_w$ can be seen from Fig.6 which indicates that the large-scale pathloss is the most influential component on $R_w$ in the long run time. It is followed by shadowing and multi-path fading components whose effect on $R_w$ is less compared to that of pathloss. The reason for this might be attributed to the model assumptions of the individual life components (pathloss has exponential distribution). However, the advantage of the Birnbaum importance metric is that it provides the reliability ranking of the individual components with respect to $R_w$ and thus helps  in identifying and eliminating reliability weaknesses.\\  

\subsection{Using Retransmissions (Redundancy)}
Retransmissions can be modeled as another instance of transmission with incremental redundancy used in techniques such as ARQ and HARQ\cite{Frenger2001a}\cite{Dahlman2013}\cite{He2010}. However, this assumes that the operating conditions are similar for both the transmission instances which doesnt hold in reality due to the time diversity (or channel coherence time) of the transmission success. This problem can be solved by adding mobility element to the RBD and unifying the time scales of the element's life distributions into milliseconds. Other techniques like Time homogeneous Markov Chains allow also allow time independency.\

\begin{figure}[h]
\centering
\includegraphics[width=0.5\textwidth]{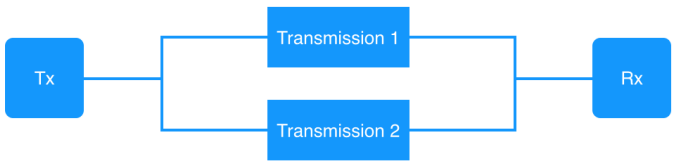}
\caption{Redundancy (Retransmission)}
\end{figure} 

Assuming that the channel is constant for the first retransmission, the two transmissions can be modeled as components connected in parallal as shown in Figure 6. The TTTF of the system can be calculated as $R_{Transmission1};\ \lor;\  R_{Transmission2}$

\begin{figure}[h!]
\centering
\includegraphics[width=0.5\textwidth]{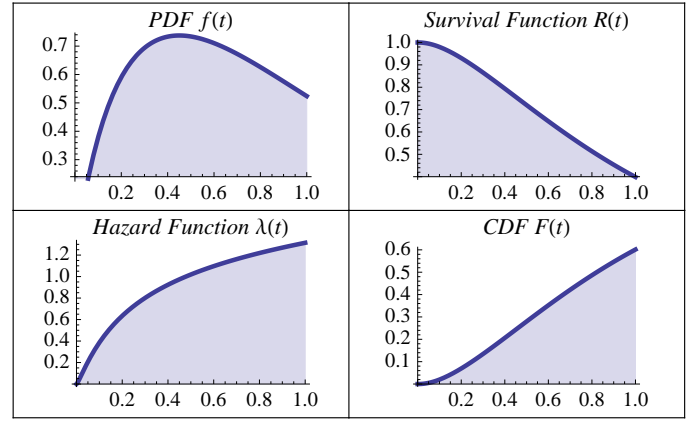}
\caption{$R_W$ KPI's with Retransmission}
\end{figure} \

Figure 7 shows the survival and hazard functions of $S_w$ with 1 retransmission. Please note the decrease in slope of the survival function with restransmission which concludes that the probability of failure is smaller than the case without retransmissions. This is also evident from figure 7 which shows the probabilities of failure before time $t$ for both the cases. 
\begin{figure}[!htb]
\minipage{0.23\textwidth}
  \includegraphics[width=\linewidth]{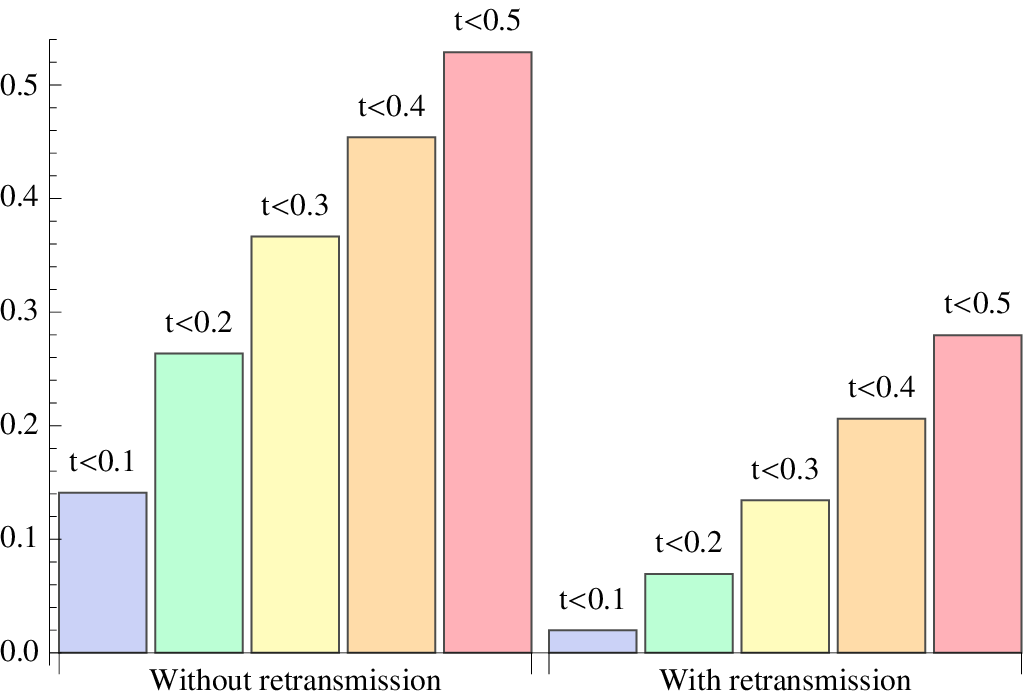}
  
\endminipage\hfill
\minipage{0.23\textwidth}
  \includegraphics[width=\linewidth]{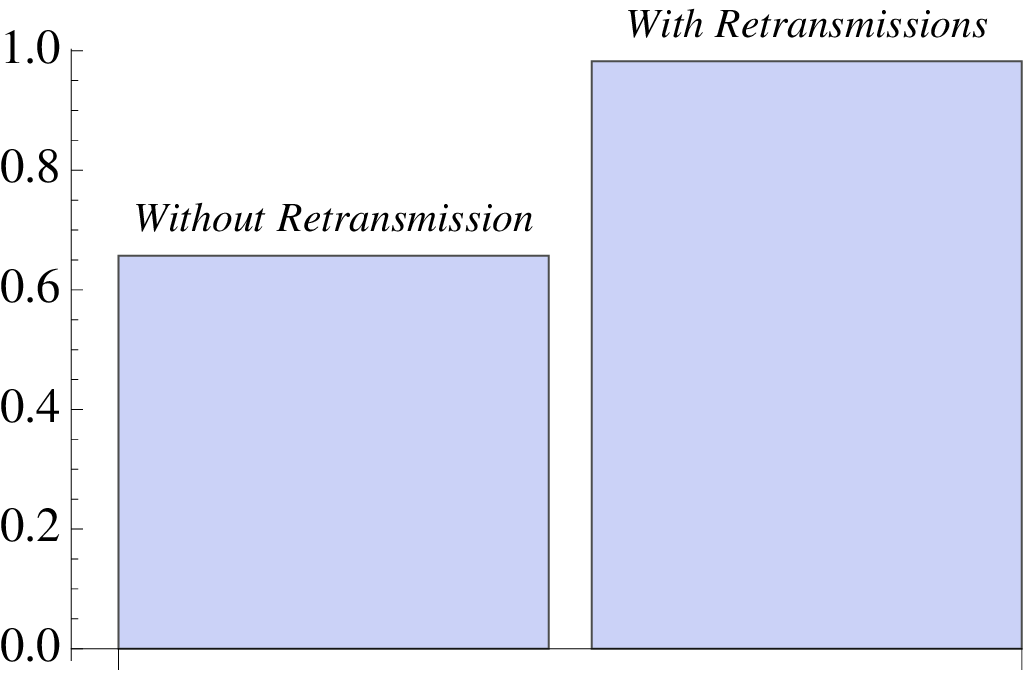}
  
\endminipage\hfill
\caption{Probabilities of failure before $t$ and Mean TTTF}
\end{figure}\

The Mean TTTF of the \textit{with-retransmission} system is calculated to be $0.98$ as compared to $0.65$ of the \textit{without-restransmission} system. 

\section{Conclusion and Future Work}
In this paper, a new framework for modeling, analyzing the reliability of wireless communication system is presented based on Reliability Analysis using RBD's. The framework is used to model and analyze the transmission success probability under the effects of channel fading and retransmissions. Reliability prediction (Availability) has been done by considering wireless channel as a \textit{non-repairable} system by means of survival $R(t)$ and hazard $\lambda(t)$ functions where Availability is the probability of the presence of a \textit{Reliable Transmission Link} (RTL) for successful transmission\cite{Sattiraju2013}. The proposed framework is cross layer compatible (reliability of the tranmission can be modeled right from the data (packet) generation till packet decoding which spans PHY, MAC, Network and Application layers). It also allows for smoother integration of the 3GPP communication systems into the legacy systems (Automotive and other Industrial systems) since it is also based on the de facto standards of the industrial reliability analysis. 

\subsection*{Future Work}
The planned Future Work can be extended in multiple dimensions as follows
\subsubsection*{Wireless Channel as a Repairable System}
In practice, a wireless transmission can be considered as a repairable system with arbitrary failure and repair rates ($\lambda$ and $\mu$). $\lambda$ is based on the factors presented in the introduction of this paper and $\mu$ is based on the \textit{pre-emptive} maintenance measures carried on the transmission (Link adaptation via power control, Modulation and Coding (MCS) schemes, Retransmissions (Redundancy) and other Error Correction techniques). In this case, Reliability of the transmission translates into Availability and Maintainability\cite{Birolini2010}

\subsubsection*{Reliability Service Composition} 		
Reliable Service Composition (RSC),i.e., the graceful degradation of the service reliability requirements has been introduced in (METIS) inorder to guarantee a certain reliability of the transmission service parameters. Hence, instead of making a binary decision "service available/not available", we can also look into the composition of the service and define composite QoS/QoE, such as one of more degraded variants of the same service.. The system reliability $R_w$ in this case can be calculated by using mult-state markov modeling.

\subsubsection*{Mobility} 
Mobility can be introduced at the Rx end by means of the physical phenomena (such as doppler shifts and channel coherence time). However, a more formal analysis of this approach is required.

\subsubsection*{Precise Channel Models} 
The considered distributions for the channel fading effects are largely emperical and does not describe the channel accurately. Precise modeling of the channel as a deterministic entity\cite{Avestimehr2011}\cite{Nuckelt2011}\cite{Schrammar2011} and precise modeling techniques such as Ray Tracing and Google Map data would more or less accurately describe the channel effects.

\bibliographystyle{IEEEtran}
\bibliography{First_paper}

\begin{thebibliography}{10}
\providecommand{\url}[1]{#1}
\csname url@samestyle\endcsname
\providecommand{\newblock}{\relax}
\providecommand{\bibinfo}[2]{#2}
\providecommand{\BIBentrySTDinterwordspacing}{\spaceskip=0pt\relax}
\providecommand{\BIBentryALTinterwordstretchfactor}{4}
\providecommand{\BIBentryALTinterwordspacing}{\spaceskip=\fontdimen2\font plus
\BIBentryALTinterwordstretchfactor\fontdimen3\font minus
  \fontdimen4\font\relax}
\providecommand{\BIBforeignlanguage}[2]{{%
\expandafter\ifx\csname l@#1\endcsname\relax
\typeout{** WARNING: IEEEtran.bst: No hyphenation pattern has been}%
\typeout{** loaded for the language `#1'. Using the pattern for}%
\typeout{** the default language instead.}%
\else
\language=\csname l@#1\endcsname
\fi
#2}}
\providecommand{\BIBdecl}{\relax}
\BIBdecl

\bibitem{Birolini2010}
A.~Birolini, \emph{{Reliability Engineering: Theory and Practice}}, 2010.

\bibitem{5222114}
W.~S. Lee, D.~Grosh, F.~Tillman, and C.~Lie, ``Fault tree analysis, methods,
  and applications 2013; a review,'' \emph{Reliability, IEEE Transactions on},
  vol. R-34, no.~3, pp. 194--203, Aug 1985.

\bibitem{Nelson2009}
\BIBentryALTinterwordspacing
W.~Nelson, \emph{Accelerated Testing: Statistical Models, Test Plans, and Data
  Analysis}, ser. Wiley Series in Probability and Statistics.\hskip 1em plus
  0.5em minus 0.4em\relax Wiley, 2009. [Online]. Available:
  \url{http://books.google.de/books?id=Dk85hUWrhp8C}
\BIBentrySTDinterwordspacing

\bibitem{Ayers2012}
\BIBentryALTinterwordspacing
M.~L. Ayers, \emph{{Telecommunications System Reliability Engineering, Theory,
  and Practice (Google eBook)}}.\hskip 1em plus 0.5em minus 0.4em\relax John
  Wiley \& Sons, 2012. [Online]. Available:
  \url{http://books.google.com/books?id=UGEPbD9gcDkC\&pgis=1}
\BIBentrySTDinterwordspacing

\bibitem{Rak2013}
J.~Rak, M.~Pickavet, K.~S. Trivedi, J.~Alonso, L.~Arie, K.~James, P.~G.~S.
  Egemen, T.~Gomes, M.~Gunkel, and K.~Walkowiak, ``{Future Research Directions
  in Design of Reliable Communication Systems},'' 2013.

\bibitem{Bai}
F.~Bai, H.~Krishnan, and V.~Sadekar, ``{Towards Characterizing and Classifying
  Communication-based Automotive Applications from a Wireless Networking
  Perspective},'' in \emph{Proceedings of IEEE Workshop on Automotive
  Networking and Applications (AutoNet)}, 2006, pp. 1--25.

\bibitem{metis2020}
\BIBentryALTinterwordspacing
``{METIS - Mobile and wireless communications Enablers for the Twenty-twenty
  Information Society}.'' [Online]. Available: \url{https://www.metis2020.com/}
\BIBentrySTDinterwordspacing

\bibitem{METIS2013}
METIS, ``{Deliverable D6.2 - Initial report on horizontal topics, results and
  system concept},'' Tech. Rep., 2013.

\bibitem{Arsal2008}
A.~Arsal, ``{A Study On Wireless Channel Models : Simulation of Fading ,
  Shadowing and Further Applications: A Thesis Submitted to in Electrical and
  Electronics Engineering by},'' Ph.D. dissertation, 2008.

\bibitem{Poikonen2009}
\BIBentryALTinterwordspacing
J.~Poikonen, ``{Efficient channel modeling methods for mobile communication
  systems},'' Ph.D. dissertation, 2009. [Online]. Available:
  \url{http://www.doria.fi/handle/10024/43085}
\BIBentrySTDinterwordspacing

\bibitem{5044933}
S.~Srinivasa and M.~Haenggi, ``Path loss exponent estimation in large wireless
  networks,'' in \emph{Information Theory and Applications Workshop, 2009}, Feb
  2009, pp. 124--129.

\bibitem{Reudink1972}
\BIBentryALTinterwordspacing
D.~Reudink, ``{Comparison of radio transmission at X-band frequencies in
  suburban and urban areas},'' \emph{IEEE Transactions on Antennas and
  Propagation}, vol.~20, no.~4, pp. 470--473, Jul. 1972. [Online]. Available:
  \url{http://ieeexplore.ieee.org/lpdocs/epic03/wrapper.htm?arnumber=1140240}
\BIBentrySTDinterwordspacing

\bibitem{okumura1968field}
Y.~Okumura, E.~Ohmori, T.~Kawano, and K.~Fukuda, ``{Field strength and its
  variability in VHF and UHF land-mobile radio service},'' \emph{Rev. Elec.
  Commun. Lab}, vol.~16, no.~9, pp. 825--873, 1968.

\bibitem{Gudmundson1991}
M.~Gudmundson, ``{Correlation model for shadow fading in mobile radio
  systems},'' \emph{Electronics Letters}, vol.~27, no.~23, p. 2145, Nov. 1991.

\bibitem{clarke1968statistical}
R.~H. Clarke, ``{A statistical theory of mobile-radio reception},'' \emph{Bell
  Syst. Tech. J}, vol.~47, no.~6, pp. 957--1000, 1968.

\bibitem{saleh1987statistical}
A.~A.~M. Saleh and R.~Valenzuela, ``{A statistical model for indoor multipath
  propagation},'' \emph{Selected Areas in Communications, IEEE Journal on},
  vol.~5, no.~2, pp. 128--137, 1987.

\bibitem{rappaport1991statistical}
T.~S. Rappaport, S.~Y. Seidel, and K.~Takamizawa, ``{Statistical channel
  impulse response models for factory and open plan building radio communicate
  system design},'' \emph{Communications, IEEE Transactions on}, vol.~39,
  no.~5, pp. 794--807, 1991.

\bibitem{Sattiraju2013}
R.~Sattiraju, H.~D. Schotten, P.~Fertl, D.~S. Gozalvez, and Z.~Ren,
  ``{Availability Indication as Key Enabler for Ultra Reliable Communication in
  5G},'' 2013.

\bibitem{Hwang2001}
\BIBentryALTinterwordspacing
F.~Hwang, ``{A new index of component importance},'' \emph{Operations Research
  Letters}, vol.~28, no.~2, pp. 75--79, Mar. 2001. [Online]. Available:
  \url{http://www.sciencedirect.com/science/article/pii/S0167637701000542}
\BIBentrySTDinterwordspacing

\bibitem{barlow1975reliability}
R.~E. Barlow, J.~B. Fussell, and N.~D. Singpurwalla, \emph{Reliability and
  fault tree analysis}.\hskip 1em plus 0.5em minus 0.4em\relax Siam
  Philadelphia, 1975, vol.~33.

\bibitem{Frenger2001a}
P.~Frenger, S.~Parkvall, and E.~Dahlman, ``{Performance comparison of HARQ with
  Chase combining and incremental redundancy for HSDPA},'' \emph{IEEE 54th
  Vehicular Technology Conference. VTC Fall 2001. Proceedings (Cat.
  No.01CH37211)}, vol.~3, 2001.

\bibitem{Dahlman2013}
\BIBentryALTinterwordspacing
E.~Dahlman, S.~Parkvall, and J.~Skold, \emph{{4G: LTE/LTE-Advanced for Mobile
  Broadband (Google eBook)}}.\hskip 1em plus 0.5em minus 0.4em\relax Academic
  Press, 2013. [Online]. Available:
  \url{http://books.google.com/books?id=AbkPAAAAQBAJ\&pgis=1}
\BIBentrySTDinterwordspacing

\bibitem{He2010}
Z.~H.~Z. He and F.~Z.~F. Zhao, ``{Performance of HARQ with AMC Schemes in LTE
  Downlink},'' \emph{Communications and Mobile Computing (CMC), 2010
  International Conference on}, vol.~2, 2010.

\bibitem{Avestimehr2011}
\BIBentryALTinterwordspacing
A.~Avestimehr, ``{Wireless network information flow: A deterministic
  approach},'' \emph{Information Theory, IEEE \ldots}, 2011. [Online].
  Available:
  \url{http://ieeexplore.ieee.org/xpls/abs\_all.jsp?arnumber=5730555}
\BIBentrySTDinterwordspacing

\bibitem{Nuckelt2011}
\BIBentryALTinterwordspacing
J.~Nuckelt, M.~Schack, and T.~K\"{u}rner, ``{Deterministic and stochastic
  channel models implemented in a physical layer simulator for Car-to-X
  communications},'' \emph{Advances in Radio Science}, vol.~9, pp. 165--171,
  Aug. 2011. [Online]. Available:
  \url{http://www.adv-radio-sci.net/9/165/2011/}
\BIBentrySTDinterwordspacing

\bibitem{Schrammar2011}
\BIBentryALTinterwordspacing
N.~Schrammar, ``{On deterministic models for wireless networks},'' 2011.
  [Online]. Available:
  \url{http://kth.diva-portal.org/smash/record.jsf?pid=diva2:409036}
\BIBentrySTDinterwordspacing

\end{thebibliography}

\section*{Acknowledgements}
Part of this work has been performed in the framework of the FP7 project ICT-317669 METIS, which is partly funded by the European Union. The authors would like to acknowledge the contributions of their colleagues in METIS, although the views expressed are those of the authors and do not necessarily represent the project.
\end{document}